\documentstyle[prl,aps,doublespace,epsf,cite,12pt]{revtex}

\title{Kondo Effect in a Quantum Antidot}

\author{M.~Kataoka, C.~J.~B.~Ford, M.~Y.~Simmons,$^{\ast}$ and
D.~A.~Ritchie}

\address{ Cavendish Laboratory, Madingley Road, Cambridge CB3 0HE, United
Kingdom }

\date{\today}

\begin{document}

\linespread{1}

\begin{spacing}{1}

\maketitle

\widetext  



\vspace{20pt}

\centerline{\bf Abstract}

\leftskip 34.8pt  

\rightskip 34.8pt

We report Kondo-like behaviour in a quantum antidot (a submicron depleted
region in a two-dimensional electron gas) in the quantum-Hall regime. When
both spin branches of the lowest Landau level encircle the antidot in a
magnetic field ($\sim 1$~T), extra resonances occur between extended edge
states via antidot bound states when tunnelling is Coulomb blockaded. These
resonances appear only in alternating Coulomb-blockaded regions, and become
suppressed when the temperature or source-drain bias is raised. Although the
exact mechanism is unknown, we believe that Kondo-like correlated tunnelling
arises from skyrmion-type edge reconstruction. This observation demonstrates
the generality of the Kondo phenomenon.

\leftskip 0pt 

\rightskip 0pt

\vspace{20pt}


One of the most well-studied many-body phenomena, the Kondo effect, arises
when an isolated electronic spin is present in a sea of free electrons. A
reason why the Kondo effect has attracted so much attention is that the same
theoretical treatment can describe many different systems, such as metals
containing magnetic impurities \cite{KONDO,HEWSON}, weakly confined quantum
dots \cite{GLAZMAN,NG,MEIR}, or even more exotic systems where the role of
spin is replaced by another internal degree of freedom \cite{COX}. Quantum
dots mimic magnetic impurities by electrostatic confinement of electrons, and
the flexibility in tuning various parameters has recently enabled the study
of Kondo phenomena in great detail
\cite{GOLDHABER,CRONENWETT,VANDERWIEL,KELLER}. Here, we report Kondo-like
correlated tunnelling in a very different system, a quantum antidot
\cite{FORD,GOLDMAN,KATAOKA,KATAOKA1}, where electrons are confined
magnetically (by the Lorentz force) around a submicron depleted region
(antidot) in a two-dimensional electron gas (2DEG). Certain features of our
results, such as the absence of spin splitting of the zero-bias anomaly,
imply that the system cannot be described by the conventional Kondo models.
We suggest that a skyrmion-type edge reconstruction around the antidot leads
to an enhancement of correlated tunnelling between extended edge states via
the antidot.

\begin{figure}

\epsfxsize=0.6\textwidth

\centerline{\epsffile{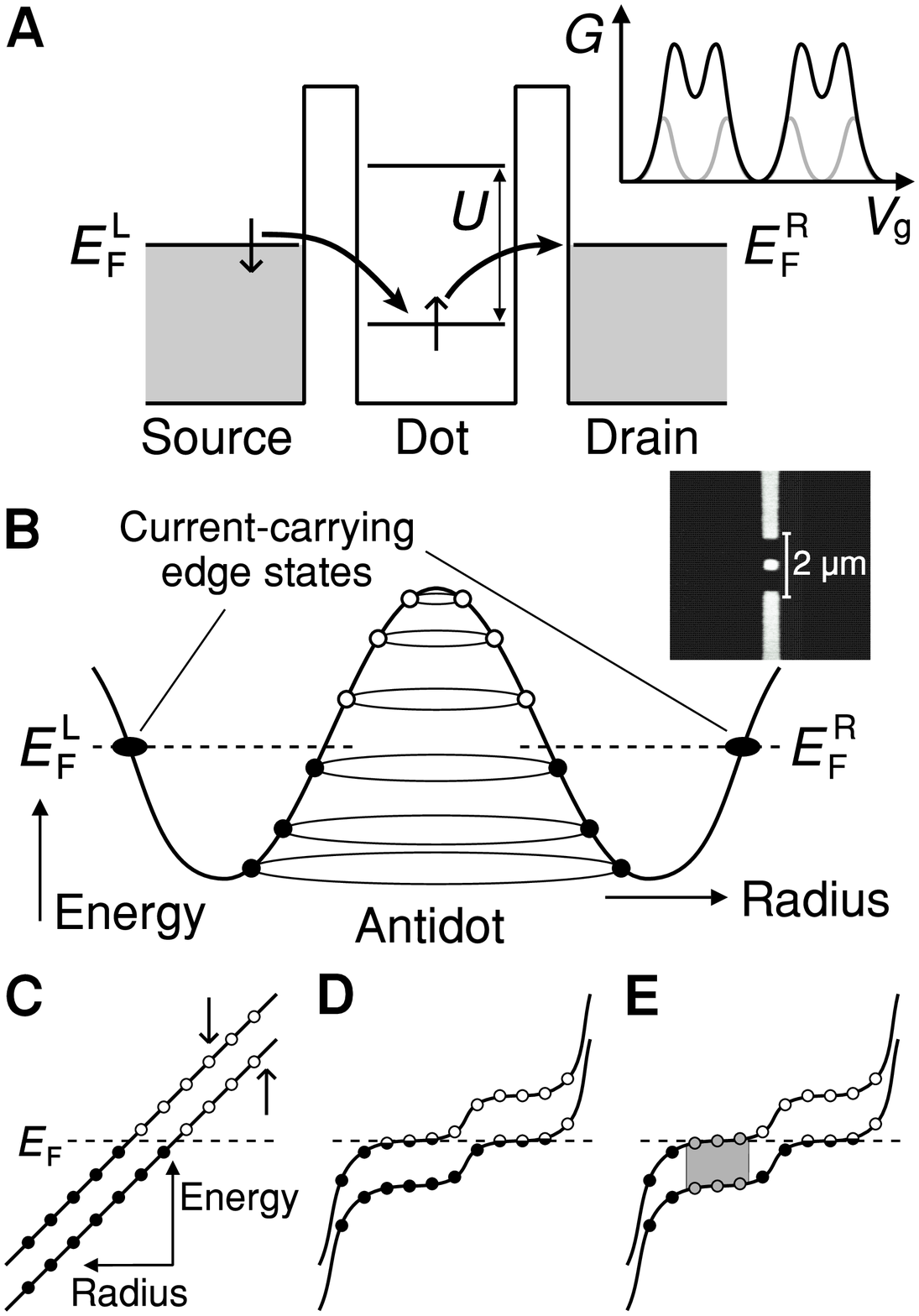}}    

\vspace{7pt}

\caption{ (A) Spin-flip tunnelling through a Coulomb-blockaded quantum dot
($U$: charging energy). The Kondo effect enhances higher-order processes of
such tunnelling at low temperature. The top-right diagram shows the
conductance $G$ of the dot versus gate voltage $V_{\rm{g}}$ with (black) and
without (grey) Kondo resonances. (B) A Landau level near the antidot and
constrictions. The circular orbits of filled and empty states are represented
by filled and open circles, respectively. Filled ovals represent the extended
edge states, which determine the current through the sample. Top-right inset:
a scanning electron micrograph of a device prior to the second-layer
metallisation.  (C) A Landau level near the Fermi energy around the antidot
with an accidental spin degeneracy between neighbouring states. (D)
Self-consistent potential with two compressible regions (one for each spin).
(E) As D but with correlated states (shadowed in grey) in the $1<\nu<2$
region. }
       
\label{fig:AD}          

\end{figure}

An impurity in a metal (or a quantum dot coupled to reservoirs) is
magnetic if it contains an unpaired electronic spin. At low temperature, the
second- and higher-order impurity scattering processes involving spin-flip of
the localised state (Fig.~\ref{fig:AD}A) is enhanced if the coupling between
the localised and delocalised electrons are antiferromagnetic. This is the
Kondo effect. As a result, in the metal, the resistivity increases as it is
cooled down, the opposite of ordinary metallic behaviour. In a quantum dot,
the transmission between reservoirs becomes enhanced. Such a Kondo resonance
occurs when the dot has an unpaired spin (i.e.\ when the number of electrons
$N$ in the quantum dot is odd).  $N$ can be changed by, for example,
varying the voltage on a gate nearby. The conductance curve as a
function of gate voltage shows pairing of resonances at low
temperature, as the Kondo effect increases the conductance of
alternating Coulomb-blockaded regions and brings the two peaks close
together (see the top-right diagram in Fig.~\ref{fig:AD}A). This is
often called odd-even behaviour.

Now, in terms of device structure, a quantum antidot (Fig.~\ref{fig:AD}B) is
quite different from a quantum dot. A magnetic field $B$ applied
perpendicular to the plane of the 2DEG quantises the kinetic energy of the
electrons into Landau levels at $(n+1/2) \hbar \omega_{\rm{c}}$, where $n$
($\ge 0$) is an integer, $\hbar = h/2\pi$ ($h$: Planck's constant), and
$\omega_{\rm{c}}$ is the cyclotron frequency. Along the bulk 2DEG boundaries,
the Landau levels rise in energy with the electrostatic potential, and
extended edge states form where the Landau levels intersect the Fermi energy.
Around the antidot, electrons form localised states, travelling
phase-coherently in closed orbits. Due to the Aharonov-Bohm effect, each
state encompasses an integer number of magnetic flux quanta $\phi_{0}=h/e$.
Therefore, the average area $S_{i}$ enclosed by the $i$th state is quantised
as $BS_{i}=i\phi_{0}$. The states are filled up to the Fermi energy (filled
circles in Fig.~\ref{fig:AD}B) and those above are empty (open circles). When
$B$ increases, each state shrinks in area (in order to keep $BS_{i}$
constant), moving towards the centre of the antidot, and accumulating a net
negative charge. Because the states are all quantised, the net charge cannot
relax until enough charge ($-e/2$) has built up, at which point an electron
leaves and the net charge jumps to $e/2$ \cite{KATAOKA}; the process then
repeats with period $\Delta B = h/eS$, where $S$ is the area enclosed by a
state at the Fermi energy. This periodic depopulation can be observed in
conductance measurements if the current-carrying extended edge states are
brought close to the antidot so that electrons tunnel between them (see
Fig.~\ref{fig:AD}B). Here, the conductance decreases on resonance, because
the resonance enhances backscattering (between opposite edges). This
resonance process when sweeping $B$ resembles that when sweeping gate voltage
in quantum dots: off resonance, tunnelling is Coulomb blockaded due to the
discreteness of the electronic charge.

Samples were fabricated from a GaAs/AlGaAs quantum-well structure containing
a 2DEG situated 300~nm below the surface with a sheet density $3 \times
10^{15}$~m$^{-2}$ and mobility 500~m$^{2}$/Vs. On top of the substrate, metal
Schottky gates (10~nm NiCr/30~nm Au) were patterned by electron-beam
lithography (top-right inset of Fig.~\ref{fig:AD}B). A second metal layer
(30~nm NiCr/70~nm Au) was patterned on top of 350~nm cross-linked
polymethylmethacrylate (PMMA) \cite{ZAILER} in order to contact the central
dot gate so that voltages can be applied to the three gates independently. A
negative voltage on the dot gate, 0.3~$\mu$m on a side, creates an antidot by
depleting electrons underneath. The depleted region can be approximated to be
circular, with a radius of $\sim 0.36 - 0.40$~$\mu$m, depending on the
voltage used; the radius is deduced from the Aharonov-Bohm period $\Delta B$.
The two side gates, which are used to bring the extended edge states close to
the antidot, form parallel one-dimensional constrictions, each with
lithographic width 0.7~$\mu$m and length 0.3~$\mu$m. The measurements were
performed in a dilution refrigerator with a base temperature $\sim 25$~mK.

\begin{figure}

\epsfxsize=\textwidth

\centerline{\epsffile{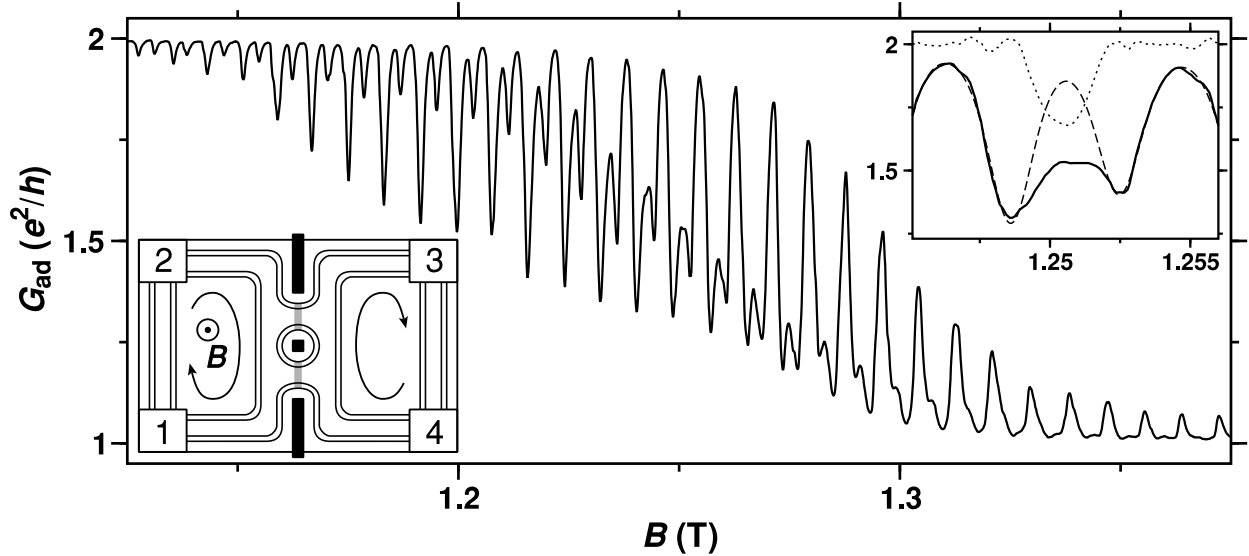}}               

\vspace{7pt}

\caption{ A typical antidot conductance $G_{\rm{ad}}$ {\em vs} $B$ curve at
25~mK between the $\nu_{\rm{c}}=2$ and 1 plateaux. The pairs of dips imply
backscattering resonances through alternating spin states. However, detailed
study of the oscillations indicates that Kondo resonances occur inside the
pairs.  Top-right inset: the results of a fit conducted for the pair around
$B=1.25$~T by using four dips proportional to the derivative of Fermi
function. The solid line is the experimental curve and the dashed line is the
fit. The fit was performed in such a way that the curves between the pairs
match as well as possible (but two dips for the pair can never fit the curve
inside the pair). The dotted line is the difference between the experimental
curve and the fit (offset by $2e^{2}/h$). Bottom-left inset: schematic
showing the sample geometry with four edge states (solid lines), two of each
spin. Arrows indicate the direction of electron flow. Grey lines show where
tunnelling occurs between the extended edge states and the antidot states.
The numbered rectangles on the corners represent ohmic contacts. }
       
\label{fig:Gad}          

\end{figure}

The conductance measurements were performed as follows. A 5~$\mu$V AC
excitation voltage at 77~Hz was applied between the ohmic contacts 1 and 3
shown in the bottom-left inset to Fig.~\ref{fig:Gad} and the current $I$ was
measured using a lock-in amplifier. Simultaneously, the diagonal voltage drop
$V_{\rm{dg}}$ between the contacts 2 and 4 was measured, and the antidot
conductance was calculated as $G_{\rm{ad}} = I/V_{\rm{dg}}$. In the quantum
Hall regime (with the correct direction of edge current flow), this
four-terminal measurement gives the `true' \cite{G2t} two-terminal
conductance $G_{\rm{ad}}=\nu_{\rm{c}}e^{2}/h$, where $\nu_{\rm{c}}$ is the
number of filled Landau levels in the antidot constrictions \cite{BUTTIKER}.
The filling factors in the constrictions and the coupling between the
extended edge states and the antidot states can be tuned by the voltages on
the side gates, and they were kept as symmetric as possible throughout the
measurements.  

The number of Landau levels forming bound states around the
antidot is defined by the number of edge states transmitted through the
constrictions.  Figure~\ref{fig:Gad} shows a typical $G_{\rm{ad}}$ {\em vs}
$B$ curve taken at 25~mK when the two spins of the lowest Landau level
encircle the antidot; $\nu_{\rm{c}}$ decreases from 2 to 1 as $B$ increases.
At low $B$ where $G_{\rm{ad}}$ is close to the $\nu_{\rm{c}}=2$ quantum-Hall
plateau value $2e^{2}/h$, a series of pairs of dips (two dips per $\Delta B$)
can be seen. As $B$ increases, and as the coupling between the leads and the
antidot becomes stronger (because the edge states move towards the centre of
the constrictions), the amplitude of the dips increases. Also, the gaps
inside pairs (intra-pair gaps) seem to be filled up, and eventually the pairs
become unrecognisable as two independent dips above 1.3~T as $G_{\rm{ad}}$
approaches the $\nu_{\rm{c}}=1$ plateau value $e^{2}/h$. Very similar
conductance curves have been observed between the $\nu_{\rm{c}}=2$ and 1
plateaux with different gate-voltage settings and with different $B$ ranging
from 0.8 to 1.5~T in many samples and on many thermal cycles. At higher
magnetic fields ($> 3$~T), the pairing of the resonances disappears, but
instead the oscillations become pure double-frequency \cite{FORD,KATAOKA1}.

These pairs may simply seem to be spin-split pairs (each dip corresponding to
a resonance of either spin). The difference in the amplitude of alternating
dips may be an indication of different coupling strength for each spin.
However, curve fitting (top-right diagram in Fig.~\ref{fig:Gad}) shows that
the feature in the intra-pair gap cannot be explained by such a simplistic
model \cite{Doublemodel}. Strikingly, the paired resonances in
Fig.~\ref{fig:Gad}, if shown upside down, look very similar to
Coulomb-blockade oscillations with Kondo resonances in quantum dots
(top-right diagram in Fig.~\ref{fig:AD}A). We argue that the discrepancy in
the intra-pair gaps is caused by Kondo resonances, based on the results from
the nonequilibrium and temperature activation measurements described below. 

\begin{figure}

\epsfxsize=\textwidth

\centerline{\epsffile{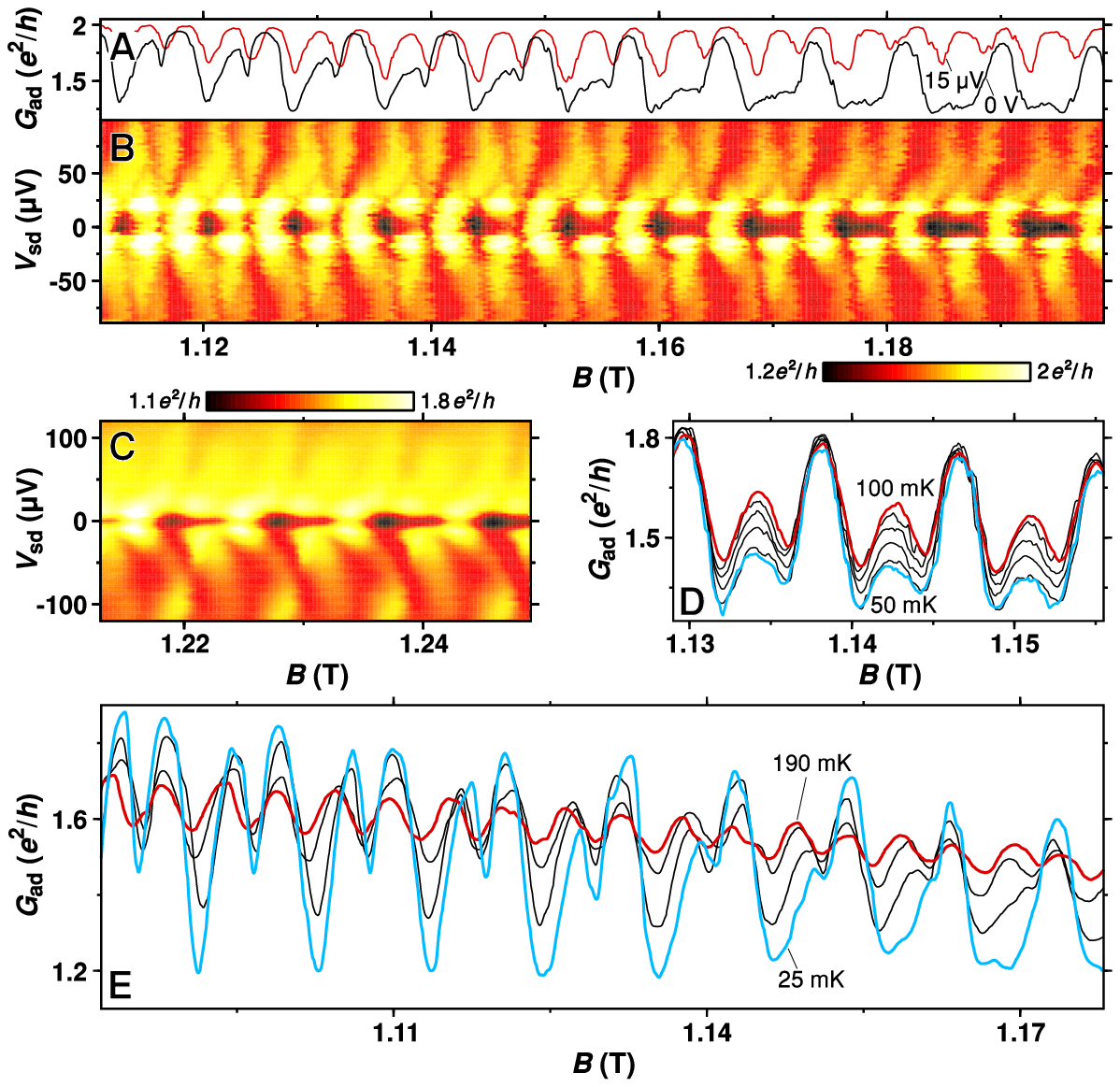}}    

\vspace{7pt}

\caption{ (A) $G_{\rm{ad}}$ {\em vs} $B$ curves at 25~mK with source-drain
bias $V_{\rm{sd}}=0$~V (black curve) and with $V_{\rm{sd}}=15$~$\mu$V (red
curve). The odd-even feature seen at zero bias almost disappears when a small
bias is applied. (B) Colour-scale plot of the differential conductance
$G_{\rm{ad}}$ against $B$ and $V_{\rm{sd}}$. As the coupling between the
leads and antidot becomes stronger (as $B$ increases), the zero-bias anomaly
becomes stronger. (C) DC-bias measurements plotted in the same manner as in
B. Here, the coupling is larger, so the diamond structure is unclear, whereas
the zero-bias anomaly is clearly visible. (D) Temperature dependence of
$G_{\rm{ad}}$ {\em vs} $B$ with small increments of temperature ($\sim
10$~mK). The feature filling the gap inside each pair weakens as the
temperature $T$ is increased, leaving the gaps inside pairs {\em better}
defined at {\em higher} temperature, which is a signature of the Kondo
effect. (E) Temperature dependence of $G_{\rm{ad}}$ {\em vs} $B$ over a wide
range of $T$. The odd-even feature at low $T$ (blue curve) disappears as $T$
is raised to 190~mK (red curve), showing almost double-frequency
oscillations. }

\label{fig:DCTdp}          

\end{figure}

The behaviour of Kondo resonances under nonequilibrium conditions has
been well studied in quantum dots
\cite{GOLDHABER,CRONENWETT,VANDERWIEL,KELLER}. Under a finite source-drain
bias, mismatch in the chemical potentials of the source and drain suppresses
the Kondo resonance. In our experiments, when a small source-drain DC bias
$V_{\rm{sd}}$ is applied in addition to the AC excitation, the `Kondo
feature' filling each intra-pair gap vanishes, leaving two well-defined dips
as shown in Fig.~\ref{fig:DCTdp}A. The Kondo features appear as horizontal
red/dark lines (zero-bias anomaly) around $V_{\rm{sd}}=0$ when $G_{\rm{ad}}$
is plotted in colour-scale against $V_{\rm{sd}}$ and $B$
(Fig.~\ref{fig:DCTdp}B). Note that here the horizontal axis is $B$ instead of
$V_{\rm{g}}$, as is normally the case for quantum dots. The zero-bias anomaly
becomes stronger as $B$, and hence the coupling, increase. The diamond-shaped
structures of red lines arise because each resonance splits into two under a
finite bias due to the difference in the chemical potentials of the two
leads. From the height of each diamond, the energy to add an extra electron
(charging energy) can be estimated to be $\sim 60$~$\mu$eV. When the coupling
is made even stronger, the zero-bias anomaly is still clearly visible,
whereas the diamond structures are smeared out due to weak confinement
(Fig.~\ref{fig:DCTdp}C).

Increasing temperature $T$ suppresses our Kondo feature, leaving the
intra-pair gaps better defined (Fig.~\ref{fig:DCTdp}D and E). At around
190~mK, the intra-pair and inter-pair gaps become almost indistinguishable.
At lower $B$ where intra-pair gaps are almost as well defined as inter-pair
gaps, increasing $T$ broadens each dip, and the conductance in both gaps
decreases (as expected without the Kondo effect). However, at higher $B$, the
conductance in the intra-pair gaps {\em increases}. Because the feature
disappears into the noise level at relatively low $T$, it was not possible to
study the temperature dependence in more detail, e.g.\ to determine the Kondo
temperature $T_{\rm{K}}$, which normally marks the crossover between
logarithmic (high $T$) and power-law (low $T$) behaviours. We note that the
amplitude of our Kondo feature decreases monotonically as $T$ increases.

The above behaviour of the antidot conductance is qualitatively very similar
to that of Kondo resonances in a quantum dot. It appears that as the antidot
states are depopulated one by one when $B$ is increased, a localised magnetic
moment arises when there is an unpaired electronic spin. Then, when the
coupling between the extended and antidot edge states is strong, Kondo
resonances enhance the tunnelling through the antidot in the
Coulomb-blockaded region. Stronger coupling results in larger $T_{\rm{K}}$,
and hence the zero-bias anomaly is more pronounced in the region closer to
the $\nu_{\rm{c}}=1$ plateau (where the wavefunctions of the extended and
antidot states strongly overlap).

However, there is a crucial difference in our results from those in quantum
dots. In a quantum dot in a comparable magnetic field, the Zeeman energy
$E_{\rm{Z}}=|g|\mu_{\rm{B}}B$ ($\mu_{\rm{B}}$ is Bohr magneton and $g=-0.44$
in bulk GaAs) usually splits the zero-bias anomaly into two parallel lines
separated by $2E_{\rm{Z}}/e$. The Kondo resonance is suppressed at
$V_{\rm{sd}}=0$ because spin degeneracy is lifted
\cite{GOLDHABER,CRONENWETT,KELLER}. At $B=1.2~T$, $E_{\rm{Z}}=30$~$\mu$eV,
but the width of our zero-bias anomaly is $\sim 20$~$\mu$V. Thus, the energy
gap between the opposite spin states must be at most 10~$\mu$eV, and probably
much smaller. This is at least a factor of 3 less than $E_{\rm Z}$.

One possible explanation for this lack of spin-splitting is to consider an
accidental degeneracy between neighbouring orbital states with opposite
spins. This might occur if $E_{\rm Z}$ happened to be close to an integer
multiple of the single-particle energy spacing $\Delta E_{\rm{sp}}$ (the
difference in potential energy between neighbouring states; see
Fig.~\ref{fig:AD}C). However, since $\Delta E_{\rm{sp}} \propto 1/B$ if a
constant potential is assumed, the ratio $E_{\rm{Z}} / \Delta E_{\rm{sp}}
\propto B^{2}$ would not stay constant over a wide range of $B$. As we have
observed the effect under many different conditions, it is highly unlikely
that this accidental degeneracy is the cause.   

Another important issue in our experiments is that the conductance dips
saturate at $e^{2}/h$ (Fig.~\ref{fig:Gad}), forming the $\nu_{\rm{c}}=1$
plateau as $B$ increases and the oscillations die away. The existence of the
$\nu_{\rm{c}}=1$ plateau implies that the extended edge states of the lower
spin state are perfectly transmitted through the constrictions, and not
coupled to the antidot. This causes another difficulty in the interpretation,
because, for ordinary Kondo resonances to occur, both spins in the leads need
to be coupled to the localised state. 

It is likely that the edge-state picture shown so far needs to be
modified to a many-body picture. Recently, we have demonstrated that a
self-consistent treatment of the antidot potential is important, especially
at higher $B$ \cite{KATAOKA1}. Taking into account the formation of two
concentric compressible rings (where states are partially filled, and are
pinned at the Fermi energy; see Fig.~\ref{fig:AD}D) \cite{CHKLOVSKII}, we
have successfully explained double-frequency Aharonov-Bohm oscillations
observed at higher $B$ \cite{FORD}. However, this is not yet enough for the
Kondo effect, as either only one spin (outer ring) would be coupled to the
leads, or, if both spins were coupled to the leads, there would be no
$\nu_{\rm c}=1$ plateau.

It has been shown that in a relatively weak magnetic field, an excited state
of a $\nu = 1$ quantum Hall liquid is not simply a spin flip, but rather, a
complicated spin texture, called a skyrmion \cite{SKYRMION}, which forms due
to strong correlations. Recent work has shown that skyrmion-type edge
reconstruction is important in quantum dots in the quantum Hall regime
\cite{HAWRYLAK,TEJEDOR}. If we assume that such correlated states form in the
regions of the antidot and extended edge states where the local filling
factor is  $1 < \nu < 2$ (Fig.~\ref{fig:AD}E), electrons with both spins
there would be allowed to tunnel, whether the $\nu \le 1$ edge states are
coupled or not. The excitation energy of correlated states is typically
smaller than $E_{\rm Z}$. If two configurations, in which the total spin
differs by one, are (almost) degenerate, this may allow spin-flip tunnelling.
It is uncertain, however, whether the odd-even behaviour can occur with such
strongly correlated states. 

In summary, we have shown that an antidot in the quantum Hall regime
qualitatively exhibits all the features of the Kondo effect, despite the lack
of an obvious reason for spin degeneracy. We have suggested possible ways in
which the edge-state picture may be modified to take account of interactions.
The fact that the antidot is an open system further complicates the problem,
in comparison with quantum dots, where only a limited number of states are
considered. A detailed theoretical calculation is needed to determine the
edge-state structure that gives rise to the Kondo-like behaviour in an
antidot.

This work was funded by the UK EPSRC. We thank I.~Smolyarenko,
N.~R.~Cooper, B.~D.~Simons, A.~S.~Sachrajda, H.-S.~Sim and V.~Falko
for useful discussions.

\vspace{5pt}

\leftline{$\ast$ Present address: School of Physics, University of New
South Wales, Sydney 2052, Australia.}

\pagebreak


\centerline{\bf Figures}

\vspace{10pt}


\pagebreak


\end{spacing}


\begin{references}
\bibitem{G2t}		`True' in the sense that the two-terminal conductance
is not affected by the series resistance of the experimental circuit.
\bibitem{Doublemodel}	In Ref.~\citen{KATAOKA1}, we gave a model with
transmission and reflection resonances in an attempt to explain similar
oscillations, however this model does not explain the temperature and
source-drain bias measurements shown here.
\end{references}
\end{document}